\documentclass[12pt,oneside]{article}
\usepackage{epsfig, cite}
\usepackage{color}
\usepackage{ifthen}
\usepackage{graphicx}

\def\p{\partial}

\def\half{{1\over 2}}
\def\({\left(}
\def\){\right)}
\def\[{\left[}
\def\]{\right]}

\def\e{\begin{equation}}
\def\q{\end{equation}}
\def\m{\begin{eqnarray}}
\def\n{\end{eqnarray}}

\begin{document}
\thispagestyle{empty} \setcounter{page}{0}

\vspace{2cm}

\begin{center}
{\huge Weak Gravity Conjecture for the Effective Field Theories with
N Species}

\vspace{1.4cm}

Qing-Guo Huang

\vspace{.2cm}

{\em School of physics, Korea Institute for
Advanced Study,} \\
{\em 207-43, Cheongryangri-Dong,
Dongdaemun-Gu, } \\
{\em Seoul 130-722, Korea}\\
\end{center}

\vspace{-.1cm}

\centerline{{\tt huangqg@kias.re.kr}} \vspace{1cm}
\centerline{ABSTRACT}
\begin{quote}%\small
\vspace{.5cm}

We conjecture an intrinsic UV cutoff for the validity of the
effective field theory with a large number of species coupled to
gravity. In four dimensions such a UV cutoff takes the form
$\Lambda=\sqrt{\lambda/ N}M_p$ for $N$ scalar fields with the same
potential $\lambda \phi_i^4$, $i=1,...,N$. This conjecture implies
that the assisted chaotic inflation or N-flation might be in the
swampland, not in the landscape. Similarly a UV cutoff
$\Lambda=gM_p/\sqrt{N}$ is conjectured for the U(1) gauge theory
with $N$ species.

\end{quote}
\baselineskip18pt

%\addtocounter{section}{1}
\noindent

\vspace{5mm}

\newpage

String theory is proposed as a candidate of quantum gravity. The
consistent perturbative superstring theory can only live in
ten-dimensional spacetime. A key problem in string theory is how it
eventually connect with experiments. To understand the effective
low-energy physics in four dimensions, string theory must be
compactified on some six dimensional manifold. Different choice of
the extra dimensional manifold leads to different low-energy
effective theory. Recent development on the flux compactification
\cite{Giddings:2001yu,Kachru:2003aw,Douglas:2006es} can help us to
stabilize some moduli. However there are so many choices of flux.
Supposing that each flux can take of order 10 values, we find of
order $10^K$ distinct solutions, where $K$ is the number of distinct
topological types of flux. Calabi-Yau manifolds typically have
$K\sim {\cal O}(10^2)$. It is quite unclear for us to get some
testable predictions for string theory because there are so many
different low-energy effective field theories.

In \cite{Vafa:2005ui} Vafa suggested that the vast series of the
metastable vacua is only semi-classically self-consistent, not
really self-consistent. We say that they are in the {\em swampland}.
The really self-consistent {\em landscape} is surrounded by the
swampland. Even though we can not pick out a unique compactification
or a unique low-energy effective field theory in four dimensions, we
still can ask which low-energy effective field theory can/cannot be
compatible with gravity.

Usually we believe that the effective field theory breaks down at
the Planck scale because the gravity becomes important and the full
theory of quantum gravity is called for. However this estimation
seems too naively. The gauge force strength is characterized by
$g^2$ and the gravitational strength is roughly $G
\Lambda^2=\Lambda^2/M_p^2$ at energy scale $\Lambda$. Gravity
becomes dominant when the energy scale goes to $gM_p$ which is much
lower than Planck scale in the perturbative region. Arkani-Hamed et
al. \cite{ArkaniHamed:2006dz} conjecture that a U(1) gauge theory
with gauge coupling $g$ has an intrinsic UV cutoff \e \Lambda \leq
gM_p \label{wog}\q for the validity. In cosmology scalar field plays
a crucial role. Both Inflaton and quintessence are scalar fields. As
a simple example, we consider the $\lambda\phi^4$ theory coupled to
gravity. The interaction strength of scale field is the
dimensionless scalar coupling $\lambda$, and the gravitational
strength is still $G\Lambda^2$. Similarly in \cite{Huang:2007gk} we
proposed a weak gravity conjecture for the $\lambda \phi^4$ theory
as follows \e \Lambda\leq \lambda^{1/2} M_p. \label{wos}\q This
conjecture infers that the single-field chaotic inflation cannot be
achieved in string landscape. Recently the authors constructed a
D-term chaotic inflation model in supergravity \cite{Kadota:2007nc}.
The potential of the scalar field takes the form $\lambda\phi^4$
which comes from the D-term of a U(1) gauge field in the same
supermultiplet. The supersymmetry implies a relationship between the
gauge coupling $g$ and the scalar coupling $\lambda$ as
$\lambda=g^2$. We see that our weak gravity conjecture for
$\lambda\phi^4$ theory is equivalent to that for the U(1) gauge
theory. For other related works on weak gravity conjecture see
\cite{Kachru:2006em,Li:2006vc,Huang:2006hc,Ooguri:2006in,Li:2006jj,Kats:2006xp,Banks:2006mm,Huang:2006pn,Huang:2006tz,Huang:2007mf,Eguchi:2007iw}.

We can imagine that there should be a lot of gauge and scalar fields
in string landscape. A loop of a light field generates a
quadratically divergent correction to the Planck scale
\cite{Dvali:2000xg,Veneziano:2001ah,ArkaniHamed:2005yv,Distler:2005hi}.
A large number of light fields can enhance radiative corrections to
the Planck mass. Demanding that the total correction $N\Lambda^2$ be
smaller than $M_p^2$ leads to an upper bound on the UV cutoff
$\Lambda$ \e \Lambda\leq \Lambda_G={M_p\over \sqrt{N}},
\label{nlp}\q where $N$ is the number of light fields. In
\cite{Dvali:2007hz} Dvali used the black hole physics to give a
strong argument to support this new energy scale for the U(1) gauge
theory with $N$ species. A large number of species of the quantum
fields implies an inevitable hierarchy between the energy scale of
the field theory and the Planck scale.

It is curious for us to ask what is the weak gravity conjecture with
a large number of field species. In this paper we will propose some
new conjectures on the U(1) gauge theory and $\lambda \phi^4$ scalar
field theory with $N$ species coupled to gravity respectively. Our
conjecture implies that even the assisted slow-roll chaotic
inflation might not be realized in the landscape. Here we only focus
on the conjectures in four dimensions, even though some of our
conjectures can be easily generalized to other dimensions.

Let's consider a scalar field theory containing $N$ scalar fields
$\phi_i$ with independent potential $V_i(\phi_i)$ for $i=1,...,N$.
For simplicity, we assume that the potential for different $\phi_i$
takes the same form \e V_i(\phi_i)=\lambda \phi_i^4. \label{potn}\q
We also assume that there are no cross-couplings between the
different scalar fields. Our results are also valid as long as the
cross-couplings can be taken as perturbations. A realization of this
field theory is that the different scalar field $\phi_i$ are stuck
to living on different points in the extra dimensions. The
cross-coupling between two fields is exponentially suppressed by the
distance between them in the extra dimensions. Let's consider the
non-supersymmetric field-theory landscape in
\cite{ArkaniHamed:2005yv}. Suppose the UV cutoff for the scalar
field theory is $\Lambda$. The distance between two fields in the
extra dimensions is required to be larger than $\Lambda^{-1}$ in
order to suppress the cross-coupling between them. This requirement
leads to a lower bound on the size of the extra dimensions $R$,
namely \e R\geq \Lambda^{-1}N^{1\over d-4}, \label{crl}\q where
$d-4$ is the number of the extra dimensions. On the other hand, the
four-dimensional Planck scale $M_p$ is related to the fundamental
Planck scale $M_d=G_d^{-{1\over d-2}}$ in $d$ dimensions by \e
M_p^2=R^{d-4}M_d^{d-2}. \label{pdp}\q Combing (\ref{pdp}) with
(\ref{crl}), we find \e M_p^2\geq {NM_d^{d-2}\over
\Lambda^{d-4}}.\label{mpd}\q Naturally the UV cutoff for the
effective field theory is lower than the fundamental Planck scale
and thus (\ref{mpd}) reads $\Lambda^2\leq M_p^2/N$ which is just the
same as (\ref{nlp}).

In the previous scenario the size of the extra dimensions is much
larger than the fundamental Planck length $M_d^{-1}$ and the length
scale $\Lambda^{-1}$ for the field theory. So we should take the
contribution of the KK modes into account. Or equivalently, for
convenience, we consider that the gravity propagates in the whole
$d$-dimensional spacetime. The action for the field theory coupled
to $d$-dimensional gravity is given by \e S={1\over 2\kappa_d^2}\int
d^dx \sqrt{-g_d} R+\int d^4x \sqrt{-g}\sum_{i=1}^{N} \left(\half (\p
\phi_i)^2+\half m^2\phi_i^2+\lambda\phi_i^4 \right), \q where
$\kappa_d^2\sim G_d$ is the Newton coupling constant in $d$
dimensions. The self-interaction strength of the scalar fields is
still characterized by the dimensionless coupling $\lambda$ and the
gravitational strength is proportional to the $d$-dimensional
Newton coupling constant $G_d\Lambda^{d-2}=(\Lambda/M_d)^{d-2}$.
Requiring that the self-interaction of the scalar fields is larger
than gravitational interaction leads to \e \Lambda\leq
\lambda^{1\over d-2}M_d.\label{etw} \q Substituting (\ref{etw}) into
(\ref{mpd}), we obtain \e \Lambda\leq \sqrt{\lambda\over
N}M_p=\lambda^\half\Lambda_G. \label{wns}\q This is the weak gravity
conjecture for the effective field theory with $N$ scalar fields and
independent potential $\lambda \phi_i^4$ for $i=1,...,N$. For $N=1$
our conjecture is just the same as that in the case with single
scalar field \cite{Huang:2007gk}.

We can easily generalize the previous arguments to U(1) gauge theory
with $N$ species in four dimensions. Similarly we also require that
these $N$ U(1) gauge fields live on the different points in the
extra dimensions and these points are separated far more than
$\Lambda^{-1}$ away from each other. Now the $N$ copies of the U(1)
gauge theory are obtained in four dimensions. The bound (\ref{crl})
on the size of extra dimensions and (\ref{mpd}) are still valid. The
strength of the gauge force is proportional to $g^2$. The gravity
also propagates in the whole $d$-dimensional spacetime. Requiring
that the gauge force be larger than the gravitational interaction
$G_d\Lambda^{d-2}=(\Lambda/M_d)^{d-2}$ yields \e \Lambda\leq
g^{2\over d-2}M_d. \q This is the weak gravity conjecture for the
U(1) gauge theory with large extra dimensions. For the detail see
\cite{Huang:2006pn} where we gave more evidence to support this
conjecture. Combing the above inequality with (\ref{mpd}), we get
the weak gravity conjecture for the U(1) gauge theory with $N$
species \e \Lambda \leq {g\over \sqrt{N}}M_p=g\Lambda_G.
\label{wng}\q This result is much more stringent than (\ref{nlp}) in
the perturbative region. Our conjecture is also consistent with the
case for the single U(1) gauge field in four dimensions
\cite{ArkaniHamed:2006dz}.

We see that $\Lambda_G=M_p/\sqrt{N}$ plays a role as the effective
gravity scale in the weak gravity conjecture for a large number of
field species. In \cite{Dvali:2007wp} the authors suggest that the
scale $\Lambda_G$ should be taken as the quantum gravity scale and
new gravitational dynamics must appear beyond this energy scale. In
\cite{Veneziano:2001ah} Veneziano argue that the dimensionless
gravity coupling $\alpha_G=G\Lambda^2$ has an upper bound $1/N$ at
one-loop level. Here we also give an argument to show that
$\Lambda_G$ is the limiting temperature for a thermal gas containing
$N$ species.

Lets consider a thermal gas with $N$ species in a box of size $L$.
This thermal gas is heated to a temperature $T$ which is assumed to
be much higher than their mass and the particles in the thermal gas
are treated as relativistic particles. The total energy of the
thermal gas is given by \e E\simeq NT^4L^3, \q and entropy takes the
form \e S\simeq NT^3L^3. \q If the size of the thermal system is
smaller than the Schwarzchild radius of the black hole with mass
$E$, this thermal system will eventually collapse into a black hole.
Requiring the thermal gas do not collapse into a black hole leads to
a maximum size of the box \e L<L_{max}={M_p\over
\sqrt{N}T^2}={\Lambda_G\over T^2}.\label{bbh} \q In
\cite{Dvali:2007wp} the authors suggested that the entropy of the
thermal gas is bounded by $M_p^2L^2$ and then we get another upper
bound on the size of the thermal system \e L<{M_p^2\over
NT^3}={\Lambda_G^2\over T^3}. \label{bws}\q But we cannot get any
conclusive results by combing (\ref{bbh}) with (\ref{bws}).  Here we
suggest to take the Bekenstein entropy bound into account. In
\cite{Bekenstein:1980jp,Bekenstein:1993dz} Bekenstein argued that
the total entropy of a system is not larger than the product of the
energy and the linear size of the system, namely \e S<S_B=EL. \q In
our case the Bekenstein entropy bound implies a lower bound on the
size of the thermal system \e L>1/T. \label{bcs}\q From the
viewpoint of quantum mechanics, the temperature can be taken as the
typical energy of a quanta in the thermal system and $1/T$ is its
wavelength which should be shorter than the size of the system.
Using (\ref{bcs}) and (\ref{bbh}), we find the temperature of the
thermal gas with $N$ species is bounded by the gravity scale
$\Lambda_G$.

From now on we switch to inflationary cosmology. The requirement
that the gravity be the weakest force leads to a stringent
constraint on inflation. In \cite{Huang:2007gk} we argued that the
single-field chaotic inflation cannot be achieved self-consistently
if we take into account the weak gravity conjecture for
$\lambda\phi^4$ theory. But the assisted chaotic inflation still
survive.

First we give a general discussion on the assisted inflation which
is governed by $N$ scalar fields with same potentials. The spacetime
geometry is quasi de Sitter space during inflation and the size of
maximum causal patch for the field theory is bounded by the Hubble
size $H^{-1}$. The Hubble parameter can be taken as the IR cutoff
for the field theory and it should be smaller than the UV cutoff
$\Lambda_G=M_p/\sqrt{N}$, namely \e N\leq {M_p^2\over H^2}\sim
S_{ds}. \label{wn}\q The number of scalar fields is bounded by the
de Sitter entropy. If the Hubble parameter exceeds the gravity
scalae $\Lambda_G$, the holographic entropy bound is violated. There
is another argument for this results in \cite{Linde:1993xx}. The
amplitude of the quantum fluctuations for the inflaton is ${H/
2\pi}$ on the Hubble scale $H^{-1}$ and thus each inflaton field
fluctuation provides a gradient energy density $(\p_\mu \delta
\phi)^2\sim (\delta \phi/H^{-1})^2\sim H^4$. The total gradient
energy density of the fluctuations of the inflatons is given by \e
N(\p_\mu \delta \phi)^2\sim NH^4.  \q A more explicit result in
\cite{Linde:1993xx} is that the quantum fluctuations of scalar
fields give a contribution to the average value of the energy
momentum tensor \e \langle T_{\mu\nu}\rangle={3NH^4\over
32\pi^2}g_{\mu\nu}. \q Requiring that ${3NH^4/(32\pi^2)}$ be not
greater than the total energy density of the inflatons $3M_p^2H^2$
yields \e N\leq {32\pi^2M_p^2\over H^2}={4\pi H^{-2}\over
G}=S_{dS};\q otherwise, inflation is shut off. The authors in
\cite{Huang:2007zt} found that the assisted chaotic inflation cannot
be eternal if we combine the weak gravity conjecture (\ref{wos})
with the entropy bound on the number of inflaton fields (\ref{wn}).

However the conjecture for the scalar field theory with $N$ species
in this paper is much more stringent than the combination of the
weak gravity conjecture for the single scalar field and the entropy
bound on the number of the fields. Here we consider the assisted
chaotic inflation. The equations of motion are given by \m
H^2=\({\dot a \over a}\)^2&=&{1\over 3M_p^2}\sum_{i=1}^{N}\(\half
{\dot \phi_i}^2+\half m^2\phi_i^2+\lambda\phi_i^4\),\\ \ddot
\phi_i+3H\dot \phi_i&=&-(m^2\phi_i+4\lambda \phi_i^3), \quad
i=1,...N.\n There is a unique attractor solution with
$\phi_1=\phi_2=...=\phi_N\equiv\phi$. Slow-roll assisted inflation
happens if the Hubble parameter evolves very slowly. It is
convenient for us to define a new slow-roll parameter for the
assisted inflation \e \epsilon_H\equiv -{\dot H\over H^2}. \q If
$\epsilon_H\ll 1$, the slow-roll assisted inflation is achieved and
the equations of motion are simplified
to be \m H^2&=&{N\over 3M_p^2}\(\half m^2\phi^2+\lambda\phi^4\),\\
3H\dot \phi&=&-(m^2\phi+4\lambda \phi^3), \quad i=1,...N, \n and the
slow-roll parameter is given by \e \epsilon_H={M_p^2\over
2N}\({dV(\phi)/d\phi\over V(\phi)}\)^2. \q Considering that the UV
cutoff $\Lambda$ in (\ref{wns}) be larger than the IR cutoff $H$, we
have \e {N\lambda\phi^4\over 3M_p^2}\leq H^2\leq \Lambda^2\leq
{\lambda \over N}M_p^2,\quad \hbox{or} \quad \phi\leq {M_p\over
\sqrt{N}}=\Lambda_G.\label{bp}\q The vacuum expectation value (VEV)
of scalar field is less than the gravity scale $\Lambda_G$. Now the
slow-roll parameter becomes \e \epsilon_H\sim {M_p^2\over
N\phi^2}\geq 1, \q which says that the slow-roll condition cannot be
achieved and the assisted chaotic inflation is in the swampland.

Even though the previous discussions are restricted to chaotic
inflation, we expect that the VEV of canonical scalar field is
bounded by $\Lambda_G$ for the case with a large number of species.
As a concrete example in string theory, the authors of
\cite{Baumann:2006cd} found the maximal variation of the canonical
inflaton field for a D3 brane in the warped background as \e \Delta
\phi=\sqrt{T_3}R\leq {2\over \sqrt{N_B}}M_p, \label{bn}\q where $R$
is the size of the throat and $N_B$ is the number of the background
D3 brane charge. If we introduce $N$ mobile D3 branes in this
scenario, the number of scalar fields is $N$ which must be bounded
by the background D3 charge $N_B$ ($N<N_B$) for the validity of the
background geometry in order to make the backreaction of the mobile
D3 branes under control. Therefore the bound in eq. (\ref{bn}) can
be written as \e \Delta \phi \leq {2\over \sqrt{N}}M_p=2\Lambda_G.
\q Ignoring the coefficient, we find that the VEV of canonical
scalar field is bounded by the gravity scale $\Lambda_G$.

To summarize, we propose the weak gravity conjectures for the
effective field theories with $N$ species coupled to gravity. The
weak gravity conjecture for the field theory with a large number of
species is more stringent than the entropy bound. If the weak
gravity conjecture is correct, it must indicate a new intrinsic
property of quantum gravity.

The conjecture for the $\lambda \phi_i^4$ theory implies that even
the assisted chaotic inflation might be in the swampland, not in the
landscape. However a possible realization of the assisted chaotic
inflation in string theory called N-flation was proposed in
\cite{Dimopoulos:2005ac}. In fact they worked in a semi-classical
way. If the conjectures in this paper are correct, N-flation is just
semi-classically self-consistent, not really self-consistent.
N-flation has the same predictions as single-field chaotic
inflation. The accuracy of PLANCK satellite which will be launched
in 2008 \cite{:2006uk} is much better than WMAP, and its
measurements will set more restricted constraints on the dynamics of
inflation. We predict that the chaotic inflation/N-flation will be
ruled out by Planck at a high confidence level, even though this
model is nicely compatible with the present WMAP data
\cite{Spergel:2006hy}. The weak gravity conjecture is still a
conjecture and its validity is uncertain, but it is a testable
conjecture. If the N-flation or chaotic inflation is confirmed by
PLANCK, it will falsify our conjecture.

In general we believe that the low-energy effective field theory is
not applicable in the over-Planckian field space. In
\cite{Ooguri:2006in} Ooguri and Vafa also gave some arguments in
string theory to support this statement. Our conjecture provides a
concrete example for it. We also propose some possible observational
consequences for the inflation models with sub-Planckian field
values in \cite{Huang:2007qz}: a lower bound on the spectral index
is obtained and the tensor perturbations can be neglected. It also
brings a constraint on the equation of state parameter of
quintessence \cite{Huang:2007mv}.

\vspace{.5cm}

\noindent {\bf Acknowledgments}

We thank S.~Kachru, K.~M.~Lee, H.~Liu and Y.~Wang for useful
discussions. We would also like to thank KITPC (Beijing) for
hospitality during a part stage of this project.


\newpage

\begin{thebibliography}{99}
\baselineskip=16pt

%\cite{Giddings:2001yu}
\bibitem{Giddings:2001yu}
  S.~B.~Giddings, S.~Kachru and J.~Polchinski,
  ``Hierarchies from fluxes in string compactifications,''
  Phys.\ Rev.\  D {\bf 66}, 106006 (2002)
  [arXiv:hep-th/0105097].
  %%CITATION = PHRVA,D66,106006;%%

%\cite{Kachru:2003aw}
\bibitem{Kachru:2003aw}
  S.~Kachru, R.~Kallosh, A.~Linde and S.~P.~Trivedi,
  ``De Sitter vacua in string theory,''
  Phys.\ Rev.\  D {\bf 68}, 046005 (2003)
  [arXiv:hep-th/0301240].
  %%CITATION = PHRVA,D68,046005;%%

%\cite{Douglas:2006es}
\bibitem{Douglas:2006es}
  M.~R.~Douglas and S.~Kachru,
  ``Flux compactification,''
  Rev.\ Mod.\ Phys.\  {\bf 79}, 733 (2007)
  [arXiv:hep-th/0610102].
  %%CITATION = RMPHA,79,733;%%

%\cite{Vafa:2005ui}
\bibitem{Vafa:2005ui}
  C.~Vafa,
  ``The string landscape and the swampland,''
  arXiv:hep-th/0509212.
  %%CITATION = HEP-TH/0509212;%%

%\cite{ArkaniHamed:2006dz}
\bibitem{ArkaniHamed:2006dz}
  N.~Arkani-Hamed, L.~Motl, A.~Nicolis and C.~Vafa,
  ``The string landscape, black holes and gravity as the weakest force,''
  JHEP {\bf 0706}, 060 (2007)
  [arXiv:hep-th/0601001].
  %%CITATION = JHEPA,0706,060;%%




%\cite{Huang:2007gk}
\bibitem{Huang:2007gk}
  Q.~G.~Huang,
  ``Weak gravity conjecture constraints on inflation,''
  JHEP {\bf 0705}, 096 (2007)
  [arXiv:hep-th/0703071].
  %%CITATION = JHEPA,0705,096;%%


%\cite{Kadota:2007nc}
\bibitem{Kadota:2007nc}
  K.~Kadota and M.~Yamaguchi,
  ``D-term chaotic inflation in supergravity,''
  Phys.\ Rev.\  D {\bf 76}, 103522 (2007)
  [arXiv:0706.2676 [hep-ph]].
  %%CITATION = PHRVA,D76,103522;%%


%\cite{Kachru:2006em}
\bibitem{Kachru:2006em}
  S.~Kachru, J.~McGreevy and P.~Svrcek,
  ``Bounds on masses of bulk fields in string compactifications,''
  JHEP {\bf 0604}, 023 (2006)
  [arXiv:hep-th/0601111].
  %%CITATION = JHEPA,0604,023;%%

%\cite{Li:2006vc}
\bibitem{Li:2006vc}
  M.~Li, W.~Song and T.~Wang,
  ``Some low dimensional evidence for the weak gravity conjecture,''
  JHEP {\bf 0603}, 094 (2006)
  [arXiv:hep-th/0601137].
  %%CITATION = JHEPA,0603,094;%%

%\cite{Huang:2006hc}
\bibitem{Huang:2006hc}
  Q.~G.~Huang, M.~Li and W.~Song,
  ``Bound on the U(1) gauge coupling in the asymptotically dS and AdS
  background,''
  JHEP {\bf 0610}, 059 (2006)
  [arXiv:hep-th/0603127].
  %%CITATION = JHEPA,0610,059;%%

%\cite{Ooguri:2006in}
\bibitem{Ooguri:2006in}
  H.~Ooguri and C.~Vafa,
  ``On the geometry of the string landscape and the swampland,''
  Nucl.\ Phys.\  B {\bf 766}, 21 (2007)
  [arXiv:hep-th/0605264].
  %%CITATION = NUPHA,B766,21;%%

%\cite{Li:2006jj}
\bibitem{Li:2006jj}
  M.~Li, W.~Song, Y.~Song and T.~Wang,
  ``A weak gravity conjecture for scalar field theories,''
  JHEP {\bf 0705}, 026 (2007)
  [arXiv:hep-th/0606011].
  %%CITATION = JHEPA,0705,026;%%

%\cite{Kats:2006xp}
\bibitem{Kats:2006xp}
  Y.~Kats, L.~Motl and M.~Padi,
  ``Higher-order corrections to mass-charge relation of extremal black holes,''
  arXiv:hep-th/0606100.
  %%CITATION = HEP-TH/0606100;%%

%\cite{Banks:2006mm}
\bibitem{Banks:2006mm}
  T.~Banks, M.~Johnson and A.~Shomer,
  ``A note on gauge theories coupled to gravity,''
  JHEP {\bf 0609}, 049 (2006)
  [arXiv:hep-th/0606277].
  %%CITATION = JHEPA,0609,049;%%

%\cite{Huang:2006pn}
\bibitem{Huang:2006pn}
  Q.~G.~Huang,
  ``Weak gravity conjecture with large extra dimensions,''
  Phys.\ Lett.\  B {\bf 658}, 155 (2008)
  [arXiv:hep-th/0610106].
  %%CITATION = HEP-TH/0610106;%%

%\cite{Huang:2006tz}
\bibitem{Huang:2006tz}
  Q.~G.~Huang and J.~H.~She,
  ``Weak gravity conjecture for noncommutative field theory,''
  JHEP {\bf 0612}, 014 (2006)
  [arXiv:hep-th/0611211].
  %%CITATION = JHEPA,0612,014;%%

%\cite{Huang:2007mf}
\bibitem{Huang:2007mf}
  Q.~G.~Huang,
  ``Gravitational correction and weak gravity conjecture,''
  JHEP {\bf 0703}, 053 (2007)
  [arXiv:hep-th/0703039].
  %%CITATION = JHEPA,0703,053;%%


%\cite{Eguchi:2007iw}
\bibitem{Eguchi:2007iw}
  T.~Eguchi and Y.~Tachikawa,
  ``Rigid Limit in N=2 Supergravity and Weak-Gravity Conjecture,''
  arXiv:0706.2114 [hep-th].
  %%CITATION = ARXIV:0706.2114;%%


%\cite{Dvali:2000xg}
\bibitem{Dvali:2000xg}
  G.~R.~Dvali and G.~Gabadadze,
  ``Gravity on a brane in infinite-volume extra space,''
  Phys.\ Rev.\  D {\bf 63}, 065007 (2001)
  [arXiv:hep-th/0008054].
  %%CITATION = PHRVA,D63,065007;%%


%\cite{Veneziano:2001ah}
\bibitem{Veneziano:2001ah}
  G.~Veneziano,
  ``Large-N bounds on, and compositeness limit of, gauge and gravitational
  interactions,''
  JHEP {\bf 0206}, 051 (2002)
  [arXiv:hep-th/0110129].
  %%CITATION = JHEPA,0206,051;%%

%\cite{ArkaniHamed:2005yv}
\bibitem{ArkaniHamed:2005yv}
  N.~Arkani-Hamed, S.~Dimopoulos and S.~Kachru,
  ``Predictive landscapes and new physics at a TeV,''
  arXiv:hep-th/0501082.
  %%CITATION = HEP-TH/0501082;%%

%\cite{Distler:2005hi}
\bibitem{Distler:2005hi}
  J.~Distler and U.~Varadarajan,
  ``Random polynomials and the friendly landscape,''
  arXiv:hep-th/0507090.
  %%CITATION = HEP-TH/0507090;%%

%\cite{Dvali:2007hz}
\bibitem{Dvali:2007hz}
  G.~Dvali,
  ``Black Holes and Large N Species Solution to the Hierarchy Problem,''
  arXiv:0706.2050 [hep-th].
  %%CITATION = ARXIV:0706.2050;%%

%\cite{Dvali:2007wp}
\bibitem{Dvali:2007wp}
  G.~Dvali and M.~Redi,
  ``Black Hole Bound on the Number of Species and Quantum Gravity at LHC,''
  arXiv:0710.4344 [hep-th].
  %%CITATION = ARXIV:0710.4344;%%


%\cite{Bekenstein:1980jp}
\bibitem{Bekenstein:1980jp}
  J.~D.~Bekenstein,
  ``A Universal Upper Bound On The Entropy To Energy Ratio For Bounded
  Systems,''
  Phys.\ Rev.\  D {\bf 23}, 287 (1981).
  %%CITATION = PHRVA,D23,287;%%

%\cite{Bekenstein:1993dz}
\bibitem{Bekenstein:1993dz}
  J.~D.~Bekenstein,
  ``Entropy Bounds And Black Hole Remnants,''
  Phys.\ Rev.\  D {\bf 49}, 1912 (1994)
  [arXiv:gr-qc/9307035].
  %%CITATION = PHRVA,D49,1912;%%


%\cite{Linde:1993xx}
\bibitem{Linde:1993xx}
  A.~D.~Linde, D.~A.~Linde and A.~Mezhlumian,
  ``From the Big Bang theory to the theory of a stationary universe,''
  Phys.\ Rev.\  D {\bf 49}, 1783 (1994)
  [arXiv:gr-qc/9306035].
  %%CITATION = PHRVA,D49,1783;%%

%\cite{Huang:2007zt}
\bibitem{Huang:2007zt}
  Q.~G.~Huang, M.~Li and Y.~Wang,
  ``Eternal Chaotic Inflation is Prohibited by Weak Gravity Conjecture,''
  JCAP {\bf 0709}, 013 (2007)
  [arXiv:0707.3471 [hep-th]].
  %%CITATION = JCAPA,0709,013;%%

%\cite{Baumann:2006cd}
\bibitem{Baumann:2006cd}
  D.~Baumann and L.~McAllister,
  ``A microscopic limit on gravitational waves from D-brane inflation,''
  Phys.\ Rev.\  D {\bf 75}, 123508 (2007)
  [arXiv:hep-th/0610285].
  %%CITATION = PHRVA,D75,123508;%%

%\cite{Dimopoulos:2005ac}
\bibitem{Dimopoulos:2005ac}
  S.~Dimopoulos, S.~Kachru, J.~McGreevy and J.~G.~Wacker,
  ``N-flation,''
  arXiv:hep-th/0507205.
  %%CITATION = HEP-TH/0507205;%%

%\cite{:2006uk}
\bibitem{:2006uk}
    [Planck Collaboration],
  ``Planck: The scientific programme,''
  arXiv:astro-ph/0604069.
  %%CITATION = ASTRO-PH/0604069;%%

%\cite{Spergel:2006hy}
\bibitem{Spergel:2006hy}
  D.~N.~Spergel {\it et al.}  [WMAP Collaboration],
  ``Wilkinson Microwave Anisotropy Probe (WMAP) three year results:
  Implications for cosmology,''
  Astrophys.\ J.\ Suppl.\  {\bf 170}, 377 (2007)
  [arXiv:astro-ph/0603449].
  %%CITATION = APJSA,170,377;%%


%\cite{Huang:2007qz}
\bibitem{Huang:2007qz}
  Q.~G.~Huang,
  ``Constraints on the spectral index for the inflation models in string
  landscape,''
  Phys.\ Rev.\  D {\bf 76}, 061303 (2007)
  [arXiv:0706.2215 [hep-th]].
  %%CITATION = PHRVA,D76,061303;%%

%\cite{Huang:2007mv}
\bibitem{Huang:2007mv}
  Q.~G.~Huang,
  ``Theoretic Limits on the Equation of State Parameter of Quintessence,''
  arXiv:0708.2760 [astro-ph].
  %%CITATION = ARXIV:0708.2760;%%



\end{thebibliography}
\end{document}